\documentclass[conference]{IEEEtran}
\IEEEoverridecommandlockouts
\usepackage{cite}
\usepackage{amsmath,amssymb,amsfonts}
\usepackage{algorithmic}
\usepackage{graphicx}
\usepackage{textcomp}
\usepackage{xcolor}
\usepackage{hyperref}

\def\BibTeX{{\rm B\kern-.05em{\sc i\kern-.025em b}\kern-.08em
    T\kern-.1667em\lower.7ex\hbox{E}\kern-.125emX}}

\begin{document}

\title{Balancing Innovation and Oversight: AI in the U.S. Treasury and IRS: A Survey}

\author{\IEEEauthorblockN{Sohail Shaikh}
\IEEEauthorblockA{\textit{Department of Computer Science} \\
\textit{George Mason University}\\
Fairfax, VA 22030 \\
mshaikh5@gmu.edu}
August, 2025
}

\maketitle
\thispagestyle{plain}
\pagestyle{plain}

\begin{abstract}
This paper explores how the U.S. Department of Treasury, particularly the Internal Revenue Service (IRS), is adopting artificial intelligence (AI) to modernize tax administration. Using publicly available information, the survey highlights the applications of AI for taxpayer support, operational efficiency, fraud detection, and audit optimization. Key initiatives include AI-powered chatbots, robotic process automation, machine learning for case selection, and advanced analytics for fraud prevention. These technologies aim to reduce errors, improve efficiency, and improve taxpayer experiences. At the same time, the IRS is implementing governance measures to ensure responsible use of AI, including privacy safeguards, transparency initiatives, and oversight mechanisms. The analysis shows that the Treasury AI strategy balances technological innovation with legal compliance, confidentiality, and public trust, reflecting a wider effort to modernize aging systems while maintaining accountability in tax collection and enforcement.
\end{abstract}

\begin{IEEEkeywords}
artificial intelligence, machine learning, tax processing, tax compliance, fraud prevention, robotic process automation, Treasury, IRS
\end{IEEEkeywords}

\section{Introduction}
The U.S. Department of the Treasury and its tax-collecting division, the Internal Revenue Service (IRS), are pursuing artificial intelligence (AI) as part of a broader effort to modernize aging IT infrastructure and applications. AI initiatives aim to improve tax collection, improve taxpayer services, prevent fraud, and reduce operational costs in addition to migrate legacy application using AI tools that "translate" old code to modernized programming languages\cite{b9}. This paper examines the AI strategies of the Treasury and IRS, focusing on both the anticipated benefits and the concerns surrounding privacy, security, and confidentiality. All information presented\footnote{Disclaimer: The information collected and presented in this survey paper is collected from public sources with no guarantee of its accuracy or lack of biases.} is drawn from publicly available sources and excludes protected or restricted knowledge, if any.

The Treasury has adopted a department-wide approach to AI implementation, designed to strengthen operations, improve public services, and ensure strong governance across its bureaus\cite{b8}. Key areas of focus include:
\begin{itemize}
    \item {AI Projects:} fraud detection and risk management, process automation, customer experience (CX) enhancements, financial market analysis, compliance-focused data analytics, and intelligent document processing.
    \item {Governance and Oversight:} establishment of an AI governance framework, risk management and accountability measures, privacy and security safeguards, human oversight mechanisms, transparency to maintain public trust, and continuous policy updates.
\end{itemize}
   
\section{AI Adoption at the IRS}
The US Department of Treasury, and specifically the Internal Revenue Service (IRS), is actively developing and deploying artificial intelligence (AI) to improve taxpayer support, operational efficiency, tax collection, and fraud prevention. Their strategy is multifaceted and evolving, with a focus on both technological innovation and taxpayer privacy. Current areas of emphasis include:

\begin{itemize}
\item Operational improvements and taxpayer support e.g., paperless processing
\item Tax collection and audit optimization
\item Fraud detection and prevention
\item AI technologies adopted, under evaluation, or planned
\item Privacy and confidentiality safeguards
\item Legacy code modernization using AI to reduced or eliminate dependence on mainframes
\end{itemize}

At present, the IRS is pursuing 68 AI-related modernization projects, including 27 dedicated to enforcement, with the goal of more efficiently identifying tax discrepancies in filings\cite{b5}.


\subsection{Operational Improvements and Taxpayer Support}

\subsubsection{AI-powered Customer Service}
The IRS has deployed AI-powered chatbots and virtual assistants to provide self-service support for routine taxpayer queries. Using natural language processing (NLP), these tools help taxpayers quickly find answers on the IRS website and phone lines, covering topics such as payment plans, transcripts, and basic account information\cite{b1, b2, b10}. By handling common inquiries instantly, they reduce wait times and free staff for more complex cases. Future plans include secure two-way messaging and expanded online account services\cite{b11}.

\subsubsection{Intelligent Automation}
The IRS employs robotic process automation (RPA) and AI-driven tools to streamline repetitive tasks such as data entry, document sorting, and taxpayer record updates. Software “robots” process forms, generate standardized letters, and manage back-office functions\cite{b3, b1}. When combined with AI, RPA supports more advanced functions, including document classification, anomaly detection, and AI-powered optical character recognition (OCR) for scanned tax forms. These systems accelerate data extraction, error checking, and case routing, enabling faster return processing and more accurate taxpayer guidance\cite{b12}. Benefits include reduced error rates, faster refunds, and improved scalability during peak filing seasons.

\subsubsection{Automated Scheduling and Call Triage}
AI systems prioritize and route incoming inquiries, schedule callbacks, and offload routine questions to virtual agents. This reduces hold times, improves response efficiency, and enables staff to focus on complex or urgent cases\cite{b1, b2}. Personalized digital tools further assist taxpayers in setting up online accounts, understanding notices, and correcting common errors.

\subsubsection{Multilingual and Accessibility Features}
AI-powered tools provide multilingual support and are being enhanced for accessibility, allowing a broader range of taxpayers to engage with IRS resources\cite{b2}.

Overall, these initiatives are expected to reduce wait times, provide faster resolutions for routine questions, enable 24/7 support, and allow IRS staff to concentrate on high-priority taxpayer needs.

\subsection{Tax Collection and Audit Optimization}

\subsubsection{Intelligent Case Selection}
The IRS uses machine learning to analyze large datasets and identify high-risk returns for audit, including those from high-income individuals, large partnerships, and hedge funds\cite{b2, b4, b5}. This targeted approach improves audit efficiency by directing resources toward cases with the greatest likelihood of non-compliance. In 2023, the Department of Treasury Office of Payment Integrity employed an improved AI-driven process to mitigate check fraud, recovering more than \$375 million\cite{b5, b12}.

\subsubsection{Enforcement and Collection}
AI models assess tax filings and financial data to detect under-reporting, improper deductions, and potential tax evasion. These models help prioritize enforcement actions based on potential returns, ensuring that resources are effectively allocated\cite{b2, b5}. For example, the Form 1040 Return Classification and Selection Models apply statistical and machine learning techniques to flag potentially non-compliant individual returns, enabling more targeted and efficient audits\cite{b18}.



\subsection{Fraud Detection and Prevention}

\subsubsection{Pattern Analysis} 
AI analyzes tax returns, bank transactions, and other datasets to detect potentially fraudulent behavior and inconsistencies in reporting \cite{b2, b5}. Treasury’s AI tools have helped prevent or recover more than \$4 billion in taxpayer losses by identifying fraudulent returns, improper payments, and check schemes. Machine learning models enable the rapid detection of high-risk transactions and the recovery of funds that would otherwise be lost \cite{b13, b14, b16}. Collaboration with the Research, Applied Analytics \& Statistics (RAAS) division enhances AI screening and prioritization of investigations, particularly for exempt organizations. Automated document-matching tools cross-check W-2s, 1099s, crypto statements, and other submissions to detect misreporting or inconsistencies. The IRS is leveraging AI to combat increasingly sophisticated fraud schemes, including those powered by emerging technologies. AI tools analyze large datasets to detect patterns and identify potentially fraudulent activity, enhancing the agency’s enforcement capabilities. While AI is effective in monitoring electronic payments, physical refund checks remain vulnerable to theft and alteration, requiring banks to implement complementary AI detection. Despite hiring 30,000 new employees to strengthen services and enforcement, AI remains essential for managing the volume and complexity of modern tax fraud\cite{b21}.

\subsubsection{Criminal Investigations} 
The IRS Criminal Investigations Branch uses AI to uncover sophisticated fraud schemes and quickly identify emerging tax evasion methods \cite{b2}. AI is used to pinpoint and track abusive tax structures and schemes designed to generate artificial deductions or credits. For instance, the agency is using AI to address non-compliance related to digital assets like cryptocurrency\cite{b19}.

In general, AI integration enables more efficient, accurate, and proactive detection and prevention of tax fraud.  



\subsection{AI Technologies Used, Explored, and Planned} 
The IRS employs artificial intelligence (AI) and machine learning to analyze large datasets, detect tax fraud, and improve compliance. AI prioritizes audits by identifying high-risk taxpayers, identifies emerging threats such as crypto-related fraud, and automates fraud detection and recovery \cite{b15}. Machine learning supports risk scoring, anomaly detection, and pattern analysis, while natural language processing (NLP) powers chatbots and extracts information from unstructured documents \cite{b3, b1}. Robotic process automation (RPA) streamlines repetitive back-office tasks, and advanced analytics guide case selection and investigations \cite{b2, b4}. Ongoing efforts focus on further integrating AI in line with technological advances and changes in the workforce \cite{b1}. The IRS leverages artificial intelligence (AI) to enhance tax compliance and reduce the tax gap (\$428 billion net gap in 2024) by improving the selection and targeting of audits. AI models are used to select representative samples of individual tax returns, helping the agency identify non-compliance trends and detect returns likely to contain errors or additional taxes owed. AI is also applied to more specialized areas. New models identify taxpayers claiming refundable credits who are at higher risk of non-compliance, outperforming previous methods in pilot studies. Additionally, AI prioritizes partnership returns for audit, allowing the IRS to focus on the highest-risk large partnerships, a complex and increasingly significant area of taxation. These AI tools collectively improve the IRS's ability to detect non-compliance and improve audit effectiveness\cite{b20}.
\subsection{Privacy and Confidentiality Safeguards} 
The IRS protects taxpayer information under Section 6103 of the Internal Revenue Code. AI initiatives are governed by the Chief Data and Analytics Officer to ensure ethical use, bias mitigation, and compliance with federal privacy requirements. Policies enforce data sharing only with authorized individuals, limit access to those with a 'need to know', require data removal or restricted re-disclosure, and include oversight of all AI projects \cite{b3}. AI systems follow the same privacy standards as other IRS technologies \cite{b6}, and transparency measures such as public dashboards help maintain accountability \cite{b7, b1}. These safeguards ensure that AI use is secure, ethical, and fully compliant.
\section{Concerns} 
Although AI promises to make tax administration more effective and fair, critics warn of risks such as bias, inaccuracy, and lack of transparency. The Government Accountability Office (GAO) and taxpayer advocates have urged the IRS to disclose more about its data sources, models, and processes \cite{b22}.  

The IRS has resisted this disclosure, citing the risk that taxpayers could 'reverse engineer' its methods to avoid compliance. Requests under the Freedom of Information Act (FOIA) have been denied, fueling concerns about accountability. These concerns intensified after a 2023 Stanford study revealed that African American taxpayers were disproportionately audited due to algorithmic bias in training data. Although the IRS has acknowledged these disparities and is developing new tools, critics argue that stronger safeguards are needed \cite{b22}.  

A further challenge is the 'black box' nature of some AI models, making it difficult for taxpayers to understand or contest audit selections. The GAO has identified insufficient documentation of IRS models, raising concerns about accountability and error correction \cite{b23}. Although the IRS states that all AI-selected cases are reviewed by a human examiner, experts caution that auditors may be reluctant to override algorithmic outputs, even when faced with legitimate statistical anomalies.  

A comprehensive AI governance strategy is needed that addresses the concerns raised by industry experts and academics, and the Treasury has taken the first step in developing and releasing its AI strategy in December 2024\cite{b8}. In March 2025, the IRS issued interim guidance on AI governance while awaiting updated directives from the White House and Treasury\cite{b24}.

Overall, while AI can strengthen tax administration, it must be deployed cautiously, with greater transparency, explainability, and oversight to maintain fairness and public trust.

In addition, budget cuts leading to AI-trained personal shortage are another concern. AI is a relatively new and specialized field and expertise is not readily available which creates a barrier to AI adoption.
\section{Conclusion}
The U.S. Department of Treasury and the IRS are leveraging artificial intelligence as a cornerstone of their modernization strategy, aiming to improve taxpayer services, optimize enforcement, and strengthen fraud detection. Early applications, including AI-powered chatbots, intelligent automation, legacy application modernization and machine learning for case selection, demonstrate measurable gains in efficiency, accuracy, and responsiveness. At the same time, governance frameworks emphasize privacy, transparency, and ethical oversight, acknowledging the sensitivity of taxpayer data and the importance of public trust. Although challenges remain to integrate AI with legacy systems and ensure fairness, the Treasury's multifaceted approach suggests a deliberate balance between innovation and accountability. As AI capabilities evolve, continued focus on privacy safeguards, policy updates, and transparency will be essential to maintain both operational improvements and public confidence in IRS mission.


\section{Future Directions}
Looking ahead, the IRS and Treasury are positioned to expand AI into emerging domains such as cryptocurrency compliance, adaptive audit strategies, and predictive analytics for tax gap reduction. Greater integration of natural language processing (NLP) tools, secure digital services, and multilingual accessibility could further improve taxpayer engagement. At the same time, advancing explainable AI and bias mitigation techniques will be critical to preserving fairness and trust. Future progress will depend not only on technological innovation but also on transparent governance and continuous public dialogue to ensure that AI strengthens, rather than undermines, confidence in tax administration.

\end{document}